\documentclass[12pt]{spieman}  
\usepackage{amsmath,amsfonts,amssymb}
\usepackage{graphicx}
\usepackage{setspace}
\usepackage{tocloft}
\usepackage{multirow}
\usepackage{soul, color, xcolor}
\soulregister\cite7
\soulregister\citeonline7
\soulregister\ref7
\usepackage{amsmath}

\title{Deep learning phase recovery: data-driven, physics-driven, or combining both?}

\author[*]{Kaiqiang Wang}
\author[*]{Edmund Y. Lam}

\affil[ ]{Department of Electrical and Electronic Engineering, The University of Hong Kong, Pokfulam,  Hong Kong SAR, China}

\renewcommand{\cftdotsep}{\cftnodots}
\cftpagenumbersoff{figure}
\cftpagenumbersoff{table} 
\begin{document} 
\maketitle

\begin{abstract}
Phase recovery, calculating the phase of a light wave from its intensity measurements, is essential for various applications, such as coherent diffraction imaging, adaptive optics, and biomedical imaging. It enables the reconstruction of an object's refractive index distribution or topography as well as the correction of imaging system aberrations. In recent years, deep learning has been proven to be highly effective in addressing phase recovery problems. Two most direct deep learning phase recovery strategies are data-driven (DD) with supervised learning mode and physics-driven (PD) with self-supervised learning mode. DD and PD achieve the same goal in different ways and lack the necessary study to reveal similarities and differences. Therefore, in this paper, we comprehensively compare these two deep learning phase recovery strategies in terms of time consumption, accuracy, generalization ability, ill-posedness adaptability, and prior capacity. What's more, we propose a co-driven (CD) strategy of combining datasets and physics for the balance of high- and low-frequency information. The codes for DD, PD, and CD are publicly available at \href{https://github.com/kqwang/DLPR}{https://github.com/kqwang/DLPR}.
\end{abstract}


{\noindent \footnotesize\textbf{*}Kaiqiang Wang,  \linkable{kqwang.optics@gmail.com}; Edmund Y. Lam,  \linkable{elam@eee.hku.hk} }

\begin{spacing}{1.25}   

\section{Introduction}
\label{sect:intro}  
Phase recovery refers to a class of methods that recover the phase of light waves from intensity measurements~\cite{wangUseDeepLearning2024}. It is active in various fields of imaging and detection, such as in bioimaging for obtaining the refractive index or thickness distribution of tissues or cells~\cite{parkQuantitativePhaseImaging2018}, in adaptive optics for characterizing aberrant wavefront~\cite{tysonPrinciplesAdaptiveOptics2022}, in coherent diffraction imaging for detecting structural information of nanomolecules~\cite{miaoExtendingMethodologyX-ray1999}, and in material inspection for measuring surface profile~\cite{leachOpticalMeasurementSurface2011b}.

Since optical detectors, such as charge-coupled device sensors, can only record the intensity/amplitude but lose the phase, one has to recover the phase from the recorded intensity indirectly. And precisely because of the loss of the phase, it is ill-posed to directly calculate the phase on the object plane from the only amplitude on the measurement plane through the forward physical model. On the one hand, the phase can be iteratively retrieved from intensity measurements with prior knowledge, i.e., phase retrieval~\cite{klibanovPhaseRetrievalProblem1995}. On the other hand, by incorporating additional information, this problem can be transformed into a well-posed one and solved directly, such as holography or interferometry with reference light~\cite{gaborNewMicroscopicPrinciple1948,goodmanIntroductionFourierOptics2017}, Shack-Hartmann wavefront sensing with micro-lens array~\cite{hartmannBermerkungen1900,shackProductionUseLenticular1971}, and the transport of intensity equation with multiple through-focus intensity images~\cite{teagueDeterministicPhaseRetrieval1983,zuoTransportIntensityEquation2020}.

\begin{figure}[!ht]
\centering
\includegraphics[width=0.6\linewidth]{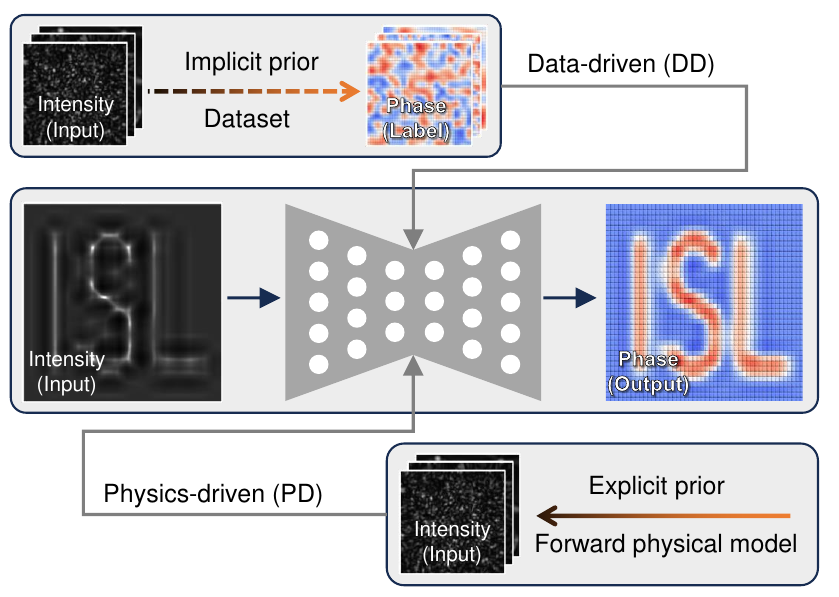}
\caption{Phase recovery network training with data-driven and physics-driven strategies.}\label{Fig:1_prin_DDandPD}
\end{figure}

In recent years, deep learning, with artificial neural networks as the carrier, has brought new solutions to phase recovery. One of the most \textit{direct} ways is to train neural networks to learn the mapping relationship from intensity measurements to the light wave phase~\cite{barbastathisUseDeepLearning2019,wangUseDeepLearning2024,wangPIERS2024}. On one hand, the training of neural networks can be driven by paired input-label datasets as an implicit prior, called data-driven (DD) strategies (see the upper part of Fig.~\ref{Fig:1_prin_DDandPD})~\cite{wangUseDeepLearning2024}. On the other hand, forward physical models can be used as an explicit prior to drive the training of neural networks with input-only datasets, called physics-driven (PD) strategies (see the lower part of Fig.~\ref{Fig:1_prin_DDandPD})~\cite{wangUseDeepLearning2024}. In addition, neural networks can also \textit{indirectly} participate in the process of phase recovery including pre-processing, in-processing (physics-connect-network, network-in-physics, and physics-in-network), and post-processing~\cite{wangUseDeepLearning2024}. Compared with classic phase recovery methods that mainly rely on physical models, deep learning methods additionally introduce prior knowledge from datasets and neural network structures to improve efficiency.

Sinha et al.~\cite{sinhaLenslessComputationalImaging2017} first demonstrated DD phase recovery with paired diffraction-phase datasets, obtained by recording diffraction images of virtual phase objects loaded on a spatial light modulator. Subsequently, DD phase recovery was successively extended to in-line holography~\cite{wangEHoloNetLearningbasedEndtoend2018}, coherent diffraction imaging~\cite{cherukaraRealtimeCoherentDiffraction2018}, Fourier ptychography~\cite{nguyenDeepLearningApproach2018}, off-axis holography~\cite{renEndtoendDeepLearning2019}, Shack-Hartmann wavefront sensing~\cite{huDeepLearningAssisted2020}, transport of intensity equation~\cite{wangTransportIntensityEquation2020}, optical diffraction tomography~\cite{pironeSpeedingReconstruction3D2022}, and electron diffractive imaging~\cite{changDeepLearningElectronDiffractive2023a}. In addition, several studies focused on more efficient neural network structures for phase recovery, such as Bayesian neural network~\cite{xueReliableDeeplearningbasedPhase2019}, generative adversarial network~\cite{liQuantitativePhaseImaging2019}, Y-Net~\cite{wangYNetOnetotwoDeep2019,wangY4NetDeepLearning2020}, residual capsule network~\cite{zengRedCapResidualEncoderdecoder2020}, recurrent neural network~\cite{huangHolographicImageReconstruction2021}, Fourier imager network~\cite{chenFourierImagerNetwork2022,chenEFINEnhancedFourier2023b}, and neural architecture search~\cite{shuNASPRNetNeuralArchitecture2022}. Some studies also used data-driven methods for pre- or post-processing of phase recovery, such as defocus distance prediction~\cite{renLearningbasedNonparametricAutofocusing2018}, resolution enhancement~\cite{renFringePatternImprovement2019}, phase unwrapping~\cite{wangOnestepRobustDeep2019}, and classification~\cite{zhuDigitalHolographicImaging2021,zhuMicroplasticPollutionMonitoring2021}.

The idea of PD phase recovery was first introduced by Boominathan et al.~\cite{boominathanPhaseRetrievalFourier} in their simulation work on Fourier ptychography. Wang et al.~\cite{wangPhaseImagingUntrained2020} first experimentally used PD to iteratively infer the phase of a phase-only object from its diffraction image directly on an untrained/initialized neural network. Afterward, it was subsequently extended to the cases of unknown defocus distances~\cite{zhangBlindNetUntrainedLearning2022}, dual wavelengths~\cite{baiDualwavelengthInlineDigital2021}, and complex-valued amplitude objects~\cite{yangDynamicCoherentDiffractive2021,yangCoherentModulationImaging2022}. In the quest for faster inference times, PD and a large number of intensity measurements were used for neural network pre-training~\cite{yangDynamicCoherentDiffractive2021,yangCoherentModulationImaging2022,bouchamaPhysicsInspiredDeepLearning2023,huangSelfsupervisedLearningHologram2023,hoidnPhysicsConstrainedUnsupervised2023}. Further, refinement of pre-trained neural networks by PD achieved higher accuracy with lower inference time~\cite{yaoAutoPhaseNNUnsupervisedPhysicsaware2022,liPhysicsenhancedNeuralNetwork2022a}. It should be noted that the PD strategies mentioned here do not include methods that use random vectors or matrices as the inputs of neural networks. For the specific differences, please refer to the italicized part on page 22 of Ref.~\citeonline{wangUseDeepLearning2024}.

DD and PD achieve the same goal in different ways and are being studied in different contexts to achieve efficient phase recovery. Therefore, it is necessary and meaningful to compare them under the same context. In this paper, we introduce the principles of DD and PD, and comparatively study them in terms of time consumption, accuracy, generalization ability, ill-posedness adaptability, and prior capacity. We also combine DD and PD as a co-driven (CD) strategy to train neural networks for high- and low-frequency information balance. What's more, to facilitate readers to get started with deep learning phase recovery quickly, we release the demonstrations of DD, PD, and CD at \href{https://github.com/kqwang/DLPR}{https://github.com/kqwang/DLPR}.

\section{Principles and Methods}

Here, we consider a classic phase recovery paradigm, recovering the phase or complex-valued amplitude of a light wave from its in-line hologram (diffraction pattern). For an object illuminated by a coherent plane wave, its hologram can be written as
\begin{equation}
 H = G(A, P) \label{eq:forward}
\end{equation}
where $H$ is the hologram, $A$ is the amplitude of light wave, $P$ is the phase of light wave, and $G(\cdot)$ is the forward propagation function, respectively. For a phase object, we assume $A=1$. Then, the purpose of phase recovery is to formulate the inverse mapping of $G(\cdot)$:
\begin{equation}
P = G^{-1}(H) \label{eq:inverse}
\end{equation}

With a supervised learning mode, DD trains neural networks with paired hologram-phase datasets $S_{H-P}=\{(H_i, P_i),i=1,\ldots, N\}$ as an \textit{implicit prior} to learn this inverse mapping~\cite{sinhaLenslessComputationalImaging2017}:

\begin{equation}
f_{\omega^{\ast}} = \mathop{\arg\min}\limits_{f_{\omega}} \sum_{i=1}^{N} \Vert f_{\omega}(H_{i})-P_{i} \Vert ^{2}_{2},\qquad \forall(H_i, P_i) \in S_{H-P} \label{con:DD}
\end{equation}
where $\Vert \cdot \Vert ^{2}_{2}$ denotes the square of the $\textit{l}_2$-norm (or other distance functions) and $f_w$ is a neural network with trainable parameters $\omega$, like weights and biases. When the optimization is complete, the trained neural network $f_{\omega^{\ast}}$ is used as an inverse mapper to infer the corresponding phase $\hat{P}_x$ from its hologram $H_x$ of an unseen object that is not in training dataset:
\begin{equation}
\hat{P}_x=f_{\omega^{\ast}}(H_x)
\end{equation}

\begin{figure}[!ht]
\centering
\includegraphics[width=0.7\linewidth]{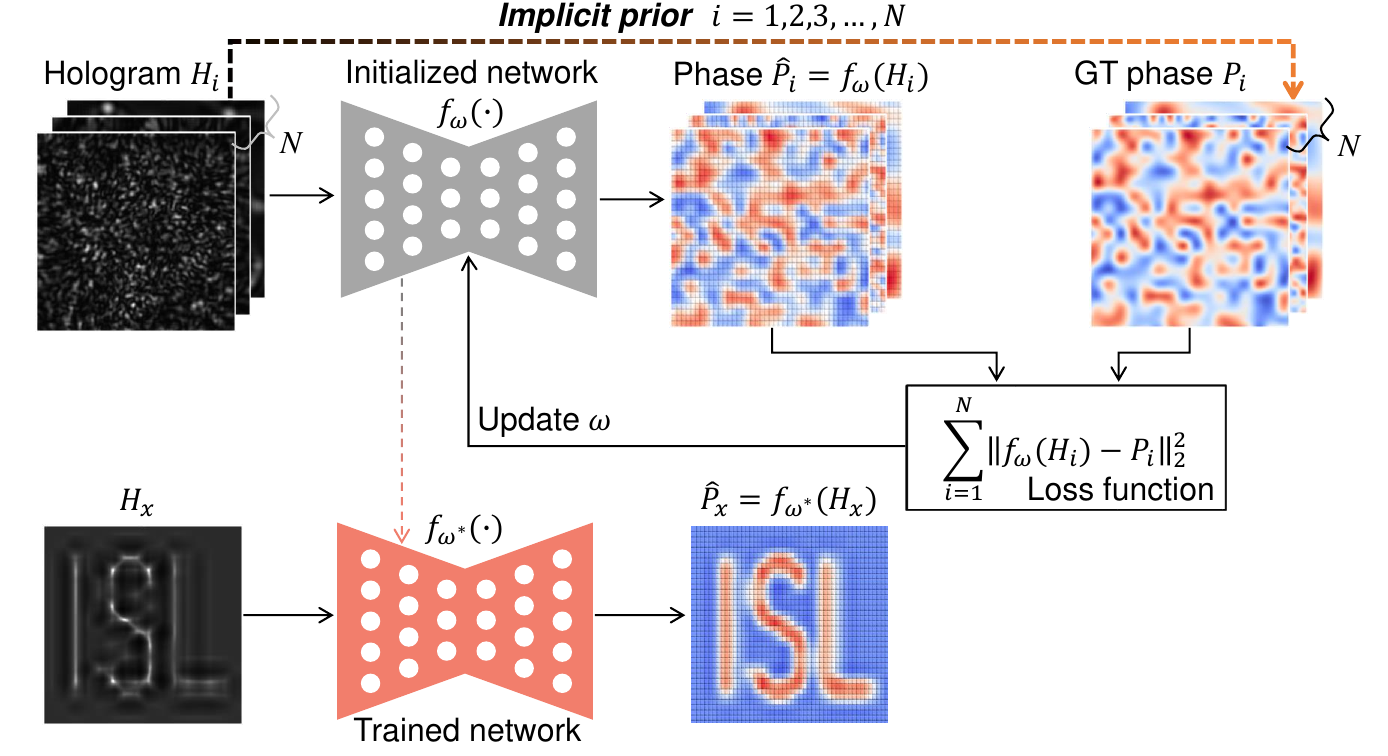}
\caption{Description of dataset-driven deep learning phase recovery methods.} \label{Fig:2_prin_DD}
\end{figure}

A visual representation of DD can be seen in Fig.~\ref{Fig:2_prin_DD}, in which holograms and phases are used as the input and ground truth (GT) of the neural network, respectively. The training dataset, collected through experiments or numerical simulations, typically contains paired data from thousands to hundreds of thousands. The training stage usually lasts for hours or even days but only takes one time. After that, the trained neural network quickly infers the phase of the unseen object after being fed its hologram. 

\begin{figure}[!ht]
\centering
\includegraphics[width=0.7\linewidth]{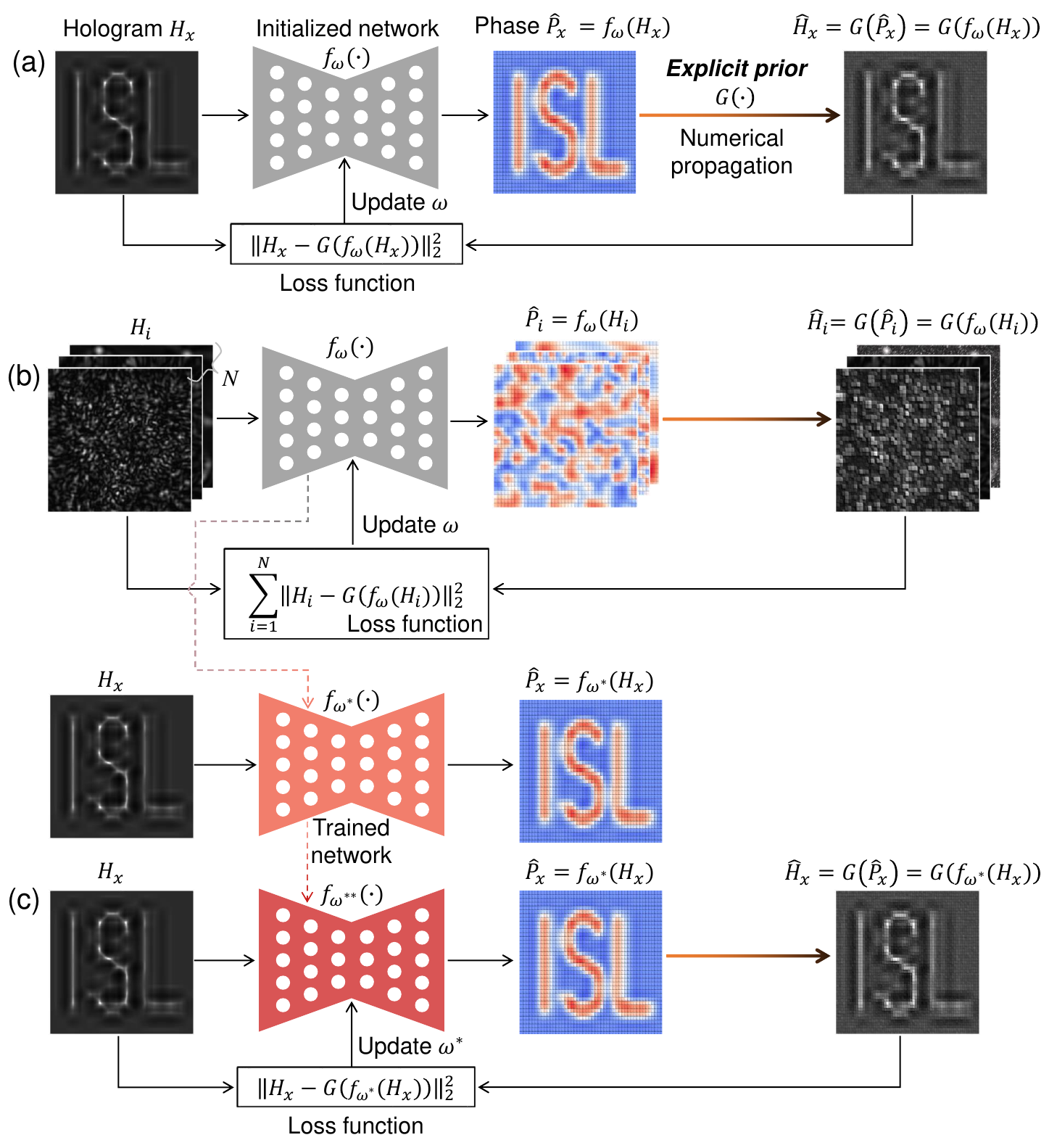}
\caption{Description of physics-driven deep learning phase recovery methods. (a) Network inference for the uPD.  (b) Network training and inference for the tPD. (c) Network training and inference for the tPDr.} \label{Fig:3_prin_PD}
\end{figure}

For physical processes that can be well modeled, such as phase recovery, PD is another available strategy. With a self-supervised learning mode, PD uses a numerical propagation $G(\cdot)$ as an \textit{explicit prior} to drive the training or inference of neural networks (Fig.~\ref{Fig:3_prin_PD}). Different from DD, which calculates the loss function in the phase domain, PD converts the network output from the phase domain to the hologram domain via numerical propagation $G(\cdot)$ and then calculates the loss function. This numerical propagation $G(\cdot)$ can be utilized to optimize the neural network in three ways: untrained PD (uPD)~\cite{wangPhaseImagingUntrained2020}, trained PD (tPD)~\cite{huangSelfsupervisedLearningHologram2023}, and tPD with refinement (tPDr)~\cite{yaoAutoPhaseNNUnsupervisedPhysicsaware2022}.

With driving of the numerical propagation $G(\cdot)$, uPD iteratively optimizes an initialized neural network $f_{\omega}(\cdot)$ to directly infer the phase $\hat{P}_x$ of an unseen object from its hologram $H_x$ (Fig.~\ref{Fig:3_prin_PD}a):
\begin{equation}
\begin{aligned}
&f_{\omega^{\ast}} = \mathop{\arg\min}\limits_{f_\omega} \Vert G(f_{\omega}(H_{x}))-H_x \Vert ^{2}_{2}\\
&\hat{P}_x = f_{\omega^{\ast}}(H_x)
\end{aligned}
\end{equation}

The most significant advantage of uPD is that it does not require any dataset to pre-process the neural network before inferences.

In tPD, the numerical propagation $G(\cdot)$ is employed to train the neural network $f_{\omega}(\cdot)$ with intensity-only training dataset $S_H=\{(H_i),i=1,\ldots, N\}$ as input, and then the trained neural network $f_{\omega^{\ast}}$ infers the phase $\hat{P}_x$ of an unseen object from its hologram $H_x$ (Fig.~\ref{Fig:3_prin_PD}b):
\begin{equation}
\begin{aligned}
&f_{\omega^{\ast}} = \mathop{\arg\min}\limits_{f_\omega} \sum_{i=1}^{N}  \Vert G(f_{\omega}(H_{i}))-H_{i} \Vert ^{2}_{2},\qquad \forall(H_i) \in S_{H} \\
&\hat{P}_x = f_{\omega^{\ast}}(H_x) \label{con:tPD}
\end{aligned}
\end{equation}

Comparing Eqs.~\ref{con:DD} and \ref{con:tPD}, we can find that the working modes of tPD and DD are similar. However, due to the use of numerical propagation $G(\cdot)$, the training dataset for tPD only requires a large number of holograms without the corresponding phase as GT.

As a strategy combining uPD and tPD, tPDr iteratively fine-tunes the tPD trained neural network $f_{\omega^{\ast}}(\cdot)$ on the hologram of the unseen object (Fig.~\ref{Fig:3_prin_PD}c):
\begin{equation}
\begin{aligned}
&f_{\omega^{\ast\ast}} = \mathop{\arg\min}\limits_{f_\omega^{\ast}} \Vert G(f_{\omega^{\ast}}(H_{x}))-H_x \Vert ^{2}_{2} \\
&\hat{P}_x = f_{\omega^{\ast\ast}}(H_x)
\end{aligned}
\end{equation}

In addition, some methods use both forward physical models and data-driven neural networks for phase recovery. On the one hand, some methods first use forward physical models to recover preliminary phases from holograms and then use data-driven neural networks to either remove unwanted components \cite{rivensonPhaseRecoveryHolographic2018,rogalskiPhysicsdrivenUniversalTwinimage2024} or perform resolution enhancement \cite{moonNoisefreeQuantitativePhase2020,chenImageEnhancementLensless2021} or convert imaging modalities\cite{wuBrightfieldHolographyCrossmodality2019}. On the other hand, some methods use data-driven neural networks to generate holograms with different propagation distances from a hologram and then recover the phase using iterative algorithms based on forward physical models\cite{luoDiffractionNetRobustSingleshot2022}. There is also an interesting way to introduce data-driven into physics-driven in the form of a generative adversarial network for phase recovery\cite{tianLenslessComputationalImaging2022}.

For the sake of clarity, we summarize DD, uPD, tPD, and tPDr according to their requirements for the physical model, the training dataset, the number of cycles needed for inference, and the learning mode in Table~\ref{tab:DDandPD}.

\begin{table}[!ht]
    \small
    \centering
    \caption{Summary of DD, uPD, tPD, and tPDr} 
    \label{tab:DDandPD}
    \begin{tabular}{cccccc}

    \hline
        \textbf{Strategy} & \textbf{Physics requirement} & \textbf{Dataset requirement} & \textbf{Inference cycles} & \textbf{Learning mode} \\ \hline
        DD & No  & Hologram-phase dataset & One time & Supervised \\ \hline
        uPD  & Numerical propagation & No & Multi times  &  self-supervised \\ \hline
        tPD & Numerical propagation & Hologram-only dataset & One time  & self-supervised  \\ \hline
        tPDr  & Numerical propagation & Hologram-only dataset & Multi times  &  self-supervised \\ \hline
    \end{tabular}
\end{table}

\section{Results and discussion}
\label{sect:results}

To avoid unnecessary distraction factors, all datasets used for comparison are generated through numerical simulation based on ImageNet, LFW, and MNIST, see Appendix A. ImageNet represents highly complex dense samples, LFW represents moderately complex dense samples, and MNIST represents simple sparse samples. Given its ubiquity in computational imaging, all methods use the same U-Net-based neural network, the specific structure of which is described in the Supplementary Material of Ref.~\citeonline{wangDeepLearningSpatial2022}. The implementation of the neural network is set uniformly, see Appendix B. The average peak signal-to-noise ratio (PSNR) and structural similarity index measure (SSIM) are used to quantify the inference accuracy. 

\subsection{Comparison of time consumption and accuracy}

In this section, ImageNet is used for dataset generation. We summarize the training settings and inference evaluation of DD, uPD, tPD, and tPDr in Table~\ref{tab:time_accuracy}.

\begin{table}[!ht]
    \centering
    \caption{Training settings and inference evaluation of DD, uPD, tPD, and tPDr}  
    \label{tab:time_accuracy} 
    
    \begin{tabular}{cccccc}
    \hline
        \textbf{Strategy} & \textbf{training datasets} & \textbf{Inference cycles} & \textbf{Inference time} & \textbf{PSNR $\uparrow$} & \textbf{SSIM $\uparrow$} \\ \hline
        DD & 10,000 pairs & 1 & ~0.02 seconds & 19.9  & 0.68 \\ \hline
        uPD & 0 &  10,000 & ~800 seconds & 25.6  & 0.94  \\ \hline
        tPD & 10,000 inputs & 1 & ~0.02 seconds  & 18.5 & 0.69 \\ \hline
        tPDr & 10,000 inputs & 1,000 & ~80 seconds & 25.1  & 0.93  \\ \hline
    \end{tabular}
\end{table}

In terms of time consumption, DD, tPD, and tPDr all require pre-training before inference, thus consuming hours or even more for neural network optimization, whereas uPD performs inference for the tested sample directly on an initialized neural network. During the inference stage of DD and tPD, the hologram of the tested sample passes through the trained neural network once in one second, while the inference process for uPD and tPDr takes several minutes for iteration. 

As for the inference accuracy, the PSNR and SSIM of DD and tPD which do quick inference once after pre-training are basically the same, and both significantly lower than uPD and tPDr which do inference multiple times. Due to the prior knowledge introduced in the pre-training stage, the initial inference of tPDr is closer to the target solution, which makes it get the same accuracy with shorter inference cycles than uPD. Specifically, with comparable accuracy, the inference time of tPDr is one-tenth that of uPD.

\begin{figure}[!ht]
\centering
\vspace{0cm}
\setlength{\abovecaptionskip}{0cm}
\includegraphics[width=0.8\linewidth]{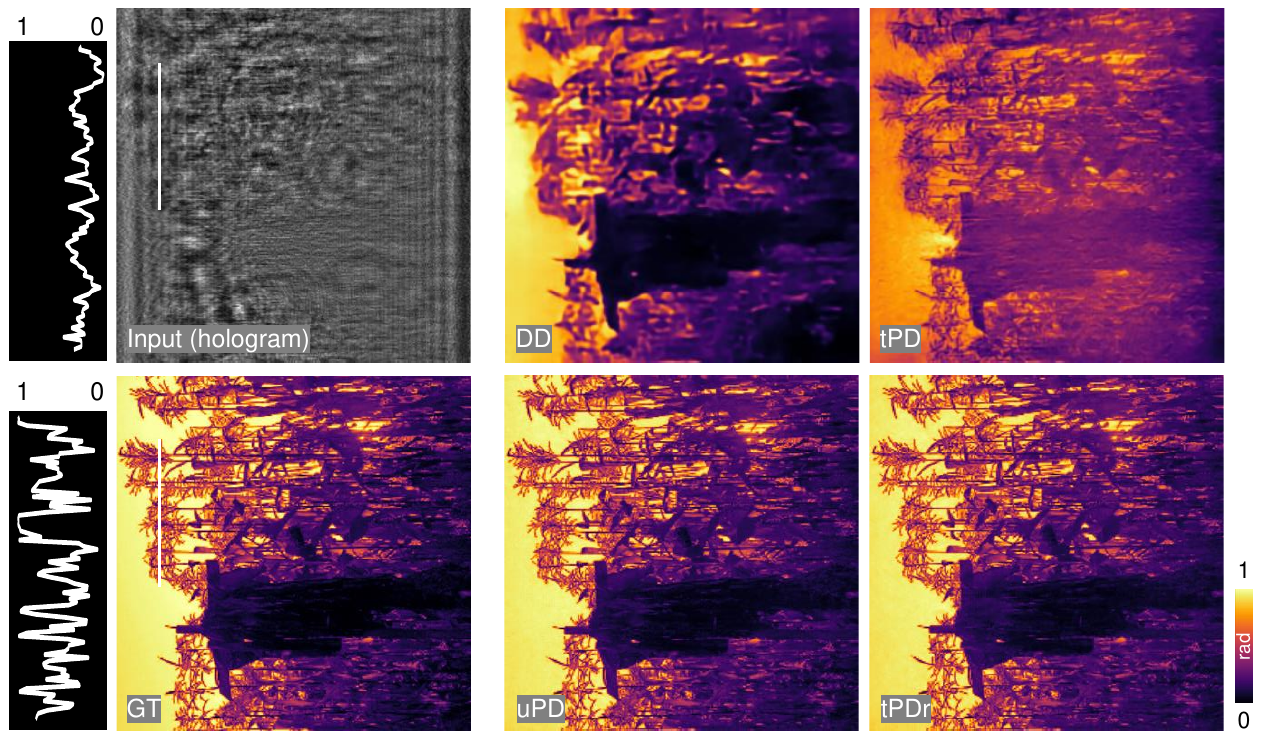}
\caption{Inference results of DD, uPD, tPD, and tPDr.}\label{Fig:4_DDandPDs}
\end{figure}

Although having the same accuracy index (Table~\ref{tab:time_accuracy}), the inference result of tPD shows better high-frequency detailed information while that of DD shows better low-frequency background information (Fig.~\ref{Fig:4_DDandPDs}). According to the frequency principle, deep neural networks are more inclined to learn low-frequency information in the data~\cite{xuFrequencyPrincipleFourier2020}. DD learns the hologram-phase mapping relationship through the loss function in the phase domain, while PD uses numerical propagation to transfer it from the phase domain to the hologram domain. On the one hand, as shown in the white curve on the left side of Fig.~\ref{Fig:4_DDandPDs}, the high-frequency phase information (steeper curve) is recorded in the diffraction fringes of the hologram which contains a more balanced high- and low-frequency information (smoother curve). This is more favorable for PD to learn those high-frequency phase information from the loss function in the hologram domain. On the other hand, the low-frequency phase causes only little contrast in the hologram, making it difficult for PD to learn low-frequency phase information, especially the plane background phase.

\begin{figure}[!ht]
\centering
\includegraphics[width=0.7\linewidth]{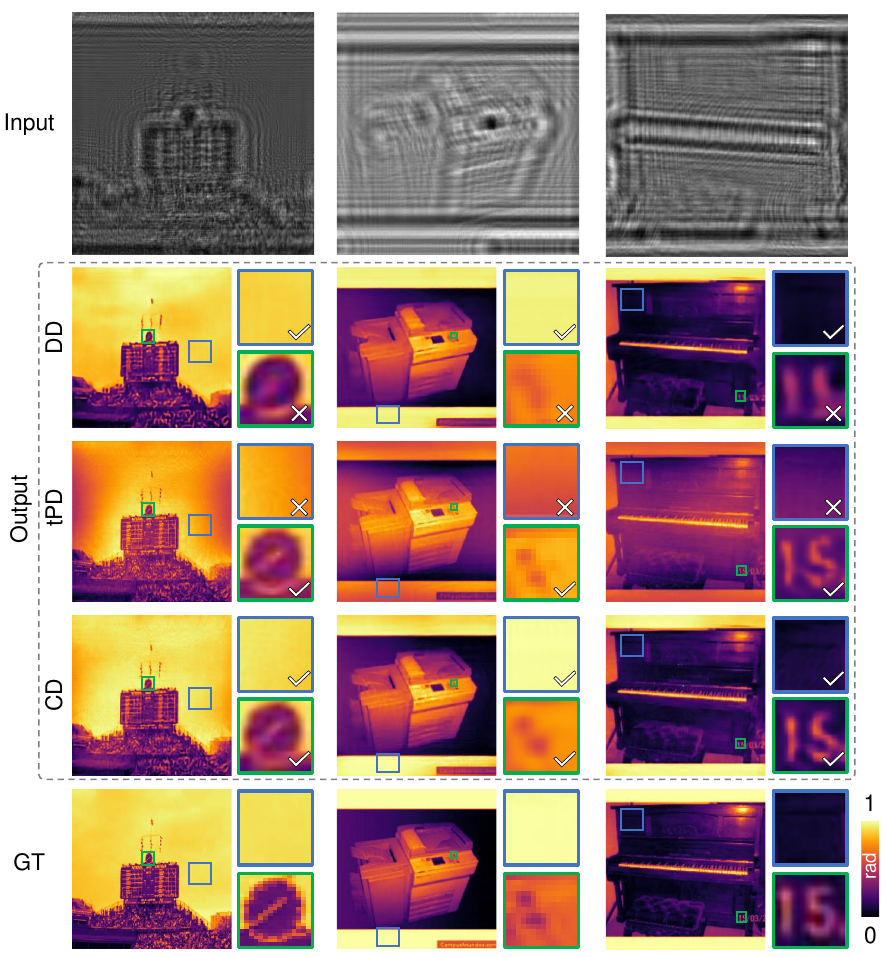}
\caption{Results of DD, tPD, and co-driven. The blue box represents low-frequency information and the green box represents high-frequency information} \label{Fig:5_DD_tPD_co_driven}
\end{figure}

In order to balance the high- and low-frequency phase information learned by the neural network, we propose to use both dataset and physics for the neural network training, named CD. The loss function of CD is derived from the weighted sum of the data-driven term and physical-driven term:
\begin{equation}
f_{\omega^{\ast}} = \mathop{\arg\min}\limits_{f_{\omega}} \sum_{i=1}^{N} \alpha \Vert f_{\omega}(H_{i})-P_{i} \Vert ^{2}_{2} + \Vert G(f_{\omega}(H_{i}))-H_{i} \Vert ^{2}_{2},\qquad \forall(H_i, P_i) \in S_{H-P}
\end{equation}
where $\alpha$ is the weight used to control the contribution of the data-driven term and physical-driven term, which is set to 0.3. As illustrated in Fig.~\ref{Fig:5_DD_tPD_co_driven}, compared to the low-frequency-tendency DD and high-frequency-tendency tPD, CD takes into account both the high-frequency phase (see the blue box) and low-frequency phase (see the green box). It should be noted that we only compared CD with DD and tPD since they all go through the neural network once for inference.

Interestingly, by comparing the inference results of holograms under different propagation distances (see Fig.~S1 of the Supplementary Material), we find that DD has a higher tolerance for defocus distance than tPD. This is most likely due to the fact that the loss function used by tPD for the neural network training is calculated in the hologram domain, and thus it is more sensitive to changes in defocus holograms than DD. In addition, CD's sensitivity to defocus distance is neutralized.

\subsection{Comparison of generalization abilit}
\begin{figure}[!ht]
\centering
\includegraphics[width=0.8\linewidth]{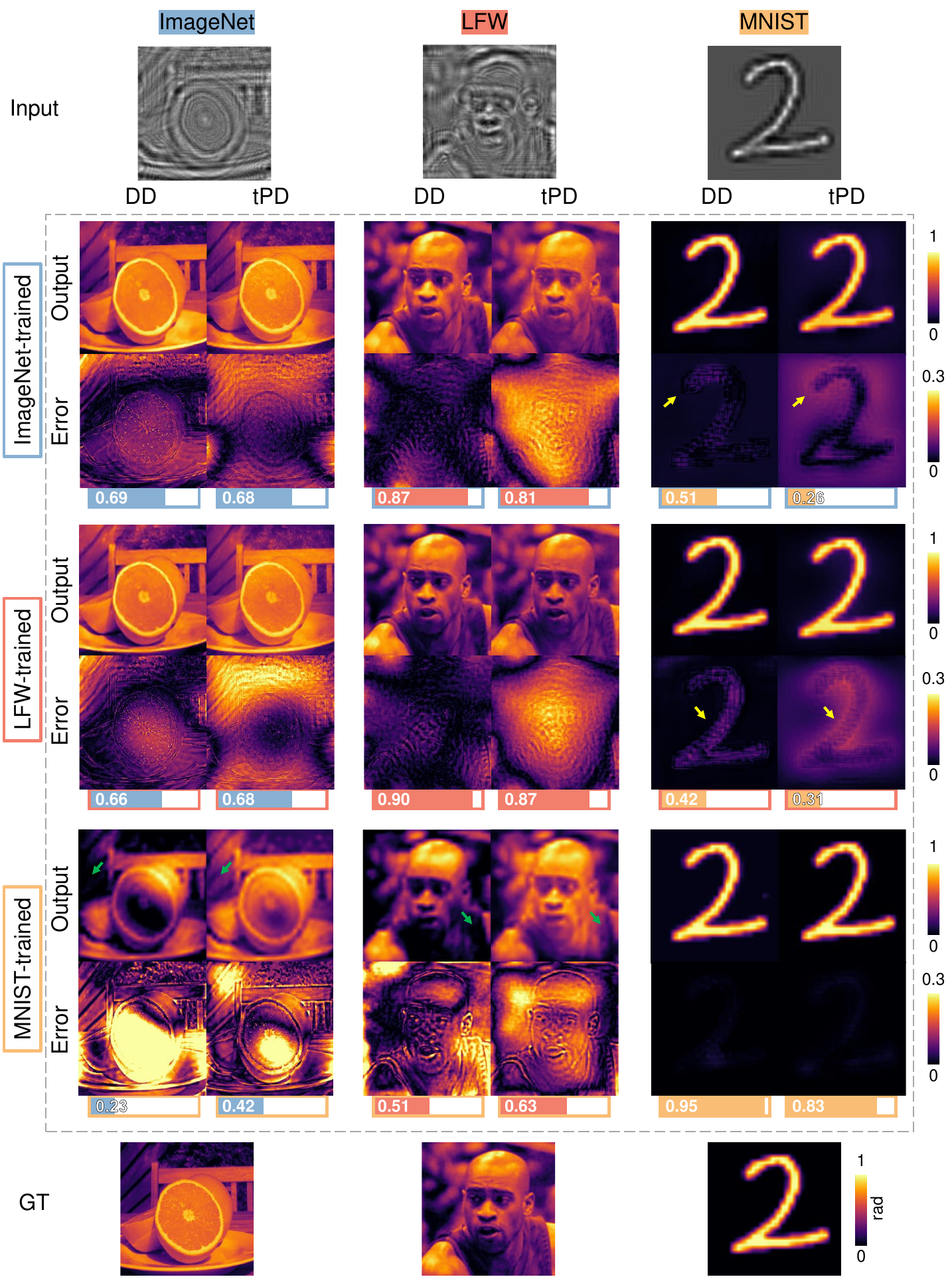}
\caption{Cross-inference results of DD and tPD for the datasets of ImageNet, LFW, and MNIST. The metric below each result is the average SSIM for that testing dataset. } \label{Fig:6_generalization}
\end{figure}

To compare the generalization ability of DD and tPD, ImageNet, LFW, and MNIST are used to generate datasets for neural network training and cross-inference, respectively. ImageNet represents dense samples, MNIST represents sparse samples, and LFW is somewhere in between. In Fig.~\ref{Fig:6_generalization}, we show the cross-inference results and their absolute error maps of a sample from ImageNet, LFW, and MNIST, and attach the average SSIM on the testing dataset below each result. 

Overall, the dataset is the main factor affecting the generalization ability of the trained neural network. Specifically, the neural networks trained by ImageNet and LFW generally perform better on all three testing datasets, while the neural networks trained by MNIST can only infer the overall distribution of ImageNet and LFW but lack detailed information. Admittedly, MNIST itself lacks detailed information, so it is reasonable that neural networks trained with it would not be able to fully infer detailed information about ImangeNet and LFW. In this extreme case, tPD is significantly better than DD, both in terms of inference results and SSIM. As can be seen in Fig.~\ref{Fig:6_generalization}, tPD infers more detailed information than DD (marked by the green arrow). Nonetheless, these results are sufficient to prove the strong generalization ability of DD and tPD, because MNIST used for training is very sparse handwritten digits with monotonous features, but the trained neural network can still do inference for the complex and feature-rich samples in ImageNet and LFW. Another thing worth noting is that for the case of using neural networks trained by ImageNet and LFW to infer MNIST, although the inference results of both tPD and DD appear to be ideal, the SSIM of tPD is much lower than that of DD. As can be seen from the absolute error maps (marked by the yellow arrow), the error in the background part of tPD is relatively larger than that of DD, which confirms a conclusion in Sec. 3.1 that tPD is not good at low-frequency phase information, especially the plane background phase.

\subsection{Comparison of ill-posedness adaptability}
Let us consider a more ill-posed case of using a neural network to simultaneously infer phase and amplitude from a hologram. In dataset generation, ImageNet, LFW, and MNIST are used to get samples containing phase and amplitude respectively, and the corresponding holograms are calculated through numerical propagation. Given that the neural network needs to output both phase and amplitude, we modified the original U-Net by paralleling another up-sampling path to build a Y-Net~\cite{wangYNetOnetotwoDeep2019}. The way tPD trains the neural network has not changed, except that there is an amplitude term in the loss function:
\begin{equation}
\begin{aligned}
&f^{P,A}_{\omega^{\ast}} = \mathop{\arg\min}\limits_{f^{P,A}_\omega} \Vert G(f^{P,A}_{\omega}(H_{x}))-H_x \Vert ^{2}_{2}\\
&\hat{P}_x, \hat{A}_x = f^{P,A}_{\omega^{\ast}}(H_x)
\end{aligned}
\end{equation}
where $f^{P,A}_\omega(\cdot)$ denotes the Y-Net that outputs phase and amplitude simultaneously. The loss function of DD is derived by weighted summation of the phase term and amplitude term:

\begin{equation}
\begin{aligned}
&f^{P,A}_{\omega^{\ast}} = \mathop{\arg\min}\limits_{f^{P,A}_{\omega}} \sum_{i=1}^{N} \Vert f^{P}_{\omega}(H_{i})-P_{i} \Vert ^{2}_{2} + \beta \Vert f^{A}_{\omega}(H_{i})-A_{i} \Vert ^{2}_{2}\\
&\hat{P}_x, \hat{A}_x = f^{P,A}_{\omega^{\ast}}(H_x)
\end{aligned}
\end{equation}
where $f^{P}_\omega(\cdot)$ and $f^{A}_\omega(\cdot)$ denote the phase path and amplitude path of Y-Net, respectively, $\beta$ is the weight used to control the contribution of the phase term and amplitude term, which is set to 0.1. 

\begin{figure}[!ht]
\centering
\includegraphics[width=0.85\linewidth]{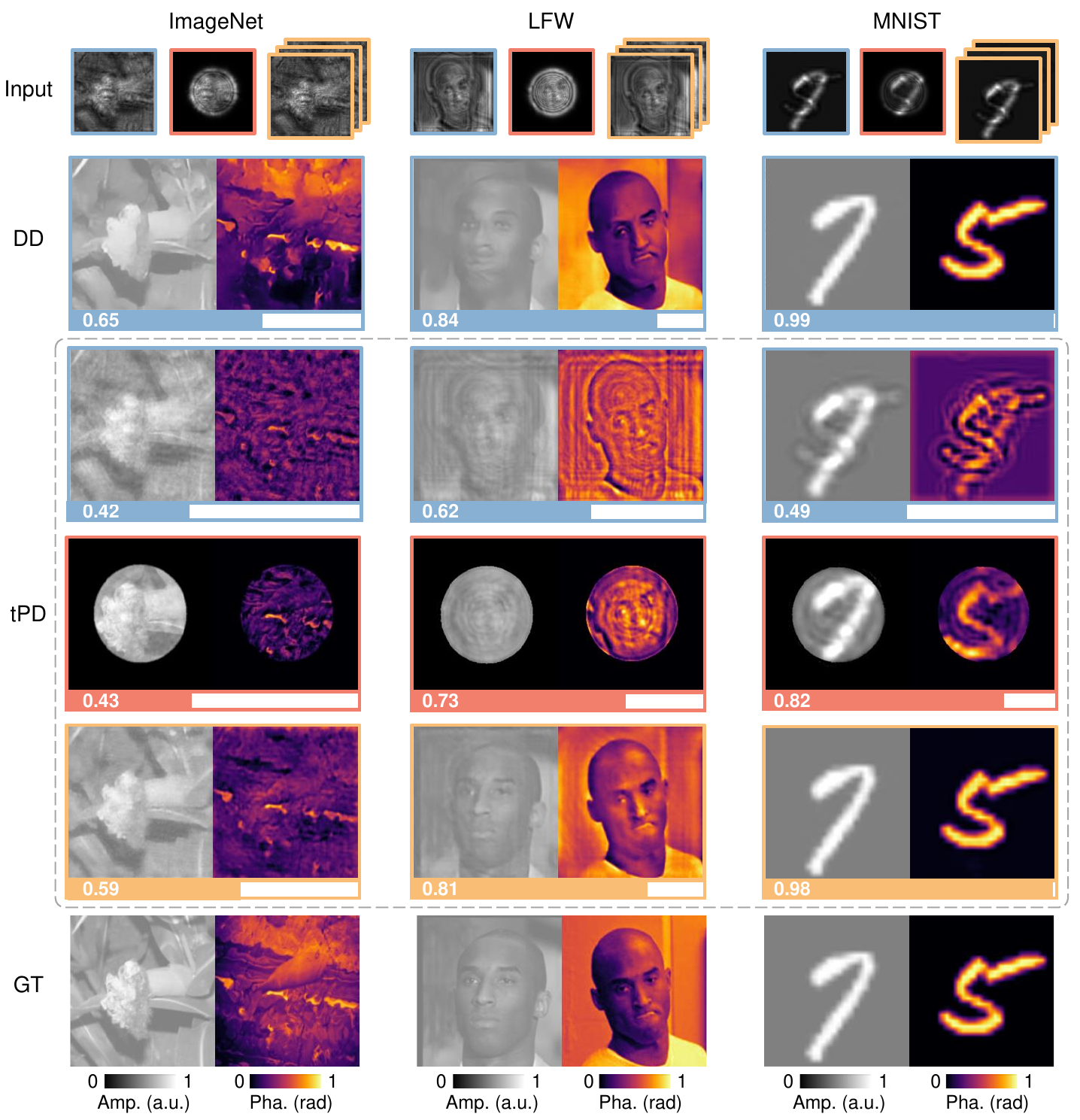}
\caption{Ill-posedness adaptability test of DD and tPD. Blue represents a single hologram as the network input, red represents a single hologram with aperture constraints as the network input, and yellow represents multiple holograms as the network input.} \label{Fig:7_ill-posedness_adaptability}
\end{figure}

The inference results of DD and tPD with single hologram input are shown in the blue part of Fig.~\ref{Fig:7_ill-posedness_adaptability}. DD can infer the phase and amplitude at the same time, because the implicit mapping relationship from holograms to phase and amplitude is completely included in the paired dataset used for the network training. As for tPD, obvious artifacts appear in the inference results and its SSIM is reduced accordingly. This means that although there are many undesirable components in the inference result, the hologram corresponding to this non-ideal phase and amplitude matches the hologram of the sample. That is, the situation of using a hologram to infer both phase and amplitude simultaneously is severely ill-posed for tPD. 

Here we show two solutions for this ill-posedness of tPD. For one thing, we introduce an aperture constraint in the sample plane to reduce the difficulty of tPD phase recovery~\cite{yangDynamicCoherentDiffractive2021}:
\begin{equation}
\begin{aligned}
&f^{P,A}_{\omega^{\ast}} = \mathop{\arg\min}\limits_{f^{P,A}_\omega} \Vert G(f^{P,A}_{\omega}(H_{x}))-H_x \Vert ^{2}_{2} + \Vert f^{A}_{\omega}(H_{x}) \cdot (1-C(r))-0_{N \times N} \Vert ^{2}_{2}\\
&\hat{P}_x, \hat{A}_x = f^{P,A}_{\omega^{\ast}}(H_x)
\end{aligned}
\end{equation}
where $C(r)$ is the aperture constraint with radius $r$ which is set to 80 pixels, and $0_{N \times N}$ denotes the zero matrix of size $N \times N$ where $N$ is set to 256. After introducing aperture constraints, the inference results of tPD for the three datasets are improved to varying degrees (see the red part of Fig.~\ref{Fig:7_ill-posedness_adaptability}). MNIST has the largest improvement, followed by LFW, and ImageNet has such limited improvement. This means that the aperture constraint works well for simple cases with less information but can hardly deal with more difficult samples. For another thing to further reduce the ill-posedness of tPD, we introduce more prior knowledge by using multiple holograms with different defocus distances as network inputs~\cite{huangSelfsupervisedLearningHologram2023}. In this case, the loss function contains three terms corresponding to different defocus distances:
\begin{equation}
\begin{split}
f^{P,A}_{\omega^{\ast}} = &\mathop{\arg\min}\limits_{f^{P,A}_\omega} \Vert G^{z_1}(f^{P,A}_{\omega}(H^{z_1}_{x}, H^{z_2}_{x}, H^{z_3}_{x}))-H^{z_1}_x \Vert ^{2}_{2}\\
& + \Vert G^{z_2}(f^{P,A}_{\omega}(H^{z_1}_{x}, H^{z_2}_{x}, H^{z_3}_{x}))-H^{z_2}_x \Vert ^{2}_{2} \\
& +\Vert G^{z_3}(f^{P,A}_{\omega}(H^{z_1}_{x}, H^{z_2}_{x}, H^{z_3}_{x}))-H^{z_3}_x \Vert ^{2}_{2} \\
\hat{P}_x, \hat{A}_x = &f^{P,A}_{\omega^{\ast}}(H^{z_1}_{x}, H^{z_2}_{x}, H^{z_3}_{x})
\end{split}
\end{equation}
where $ G^{z_1}(\cdot), G^{z_2}(\cdot), G^{z_3}(\cdot)$ donate the numerical propagation of different distances, and $ H^{z_1}_{x}, H^{z_2}_{x}, H^{z_3}_{x}$ donate holograms with different defocus distances, where $z_1,z_2,z_3$ are set to 20mm, 40mm, and 60mm respectively. Compared to a single hologram input, two more holograms introduce sufficient prior knowledge for tPD, resulting in a significant improvement in the trained neural network, both for the simple MNIST and the complex LFW and ImageNet (see the yellow part of Fig.~\ref{Fig:7_ill-posedness_adaptability}).

\subsection{Comparison of prior capacity}
\begin{figure}[!ht]
\centering
\includegraphics[width=0.75\linewidth]{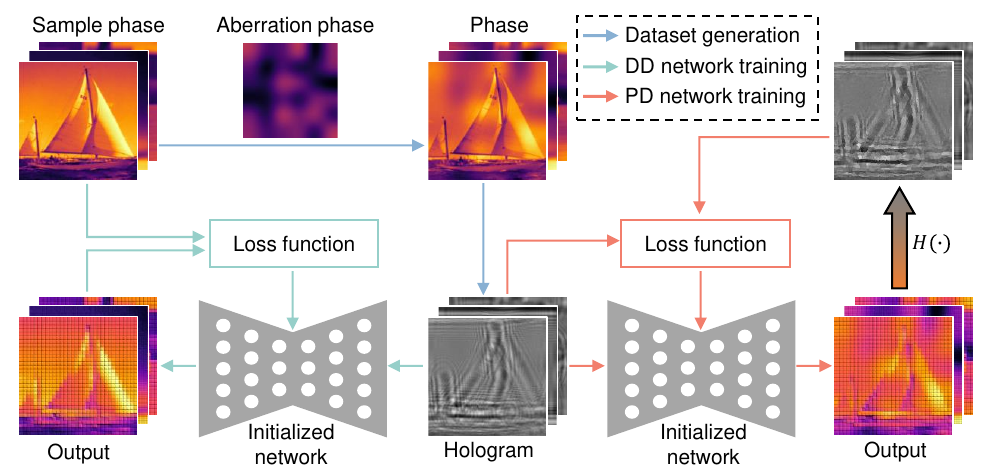}
\caption{Dataset generation and network training for the case of imaging aberration} \label{Fig:8_prior_capacity_network_train}
\end{figure}

tPD uses numerical propagation as an explicit prior to train the neural network, so the neural network learns priors from numerical propagation. DD trains a neural network with paired datasets, which means that the neural network learns all implicit priors contained in the dataset even if it is outside the numerical propagation.  For example, in the presence of imaging aberration, there will be both sample and aberration information in the hologram. Here, we use ImageNet as the sample phase and a random phase generated by the random matrix enlargement (RME)~\cite{wangOnestepRobustDeep2019,wangDeepLearningSpatial2022} as the aberration phase to generate a dataset for the comparison of DD and tPD. The process of dataset generation and network training is shown in Fig.~\ref{Fig:8_prior_capacity_network_train}, where blue represents the dataset generation part, green represents the network training part of DD, and red represents the network training part of tPD.

\begin{figure}[ht!]
\centering
\includegraphics[width=0.6\linewidth]{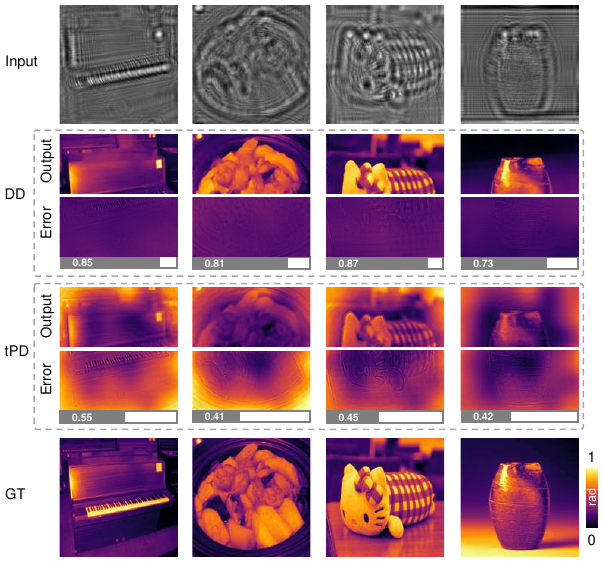}
\caption{Prior capacity test of DD and tPD} \label{Fig:9_prior_capacity}
\end{figure}

We illuminate the inference results and absolute error maps of four samples in Fig.~\ref{Fig:9_prior_capacity}. As expected, DD infers the sample phase while removing the imaging aberration phase, while the inference result of tPD includes both the sample phase and the aberration phase. Accordingly, the SSIM of DD is much higher than that of tPD. In DD, the hologram contains unwanted aberration information, but the ground truth only contains sample information, which means that the dataset implicitly contains both the prior for phase recovery and the prior for aberration removal. As for tPD, the prior for the network training is derived from numerical propagation, which allows both the sample information and the aberration information in the hologram to be recovered. It should be noted that the results of uPD also contain the unwanted aberration phase just like that of tPD.

\subsection{Comparison of experimental data}

\begin{figure}[!ht]
\centering
\vspace{0cm}
\setlength{\abovecaptionskip}{0cm}
\includegraphics[width=0.85\linewidth]{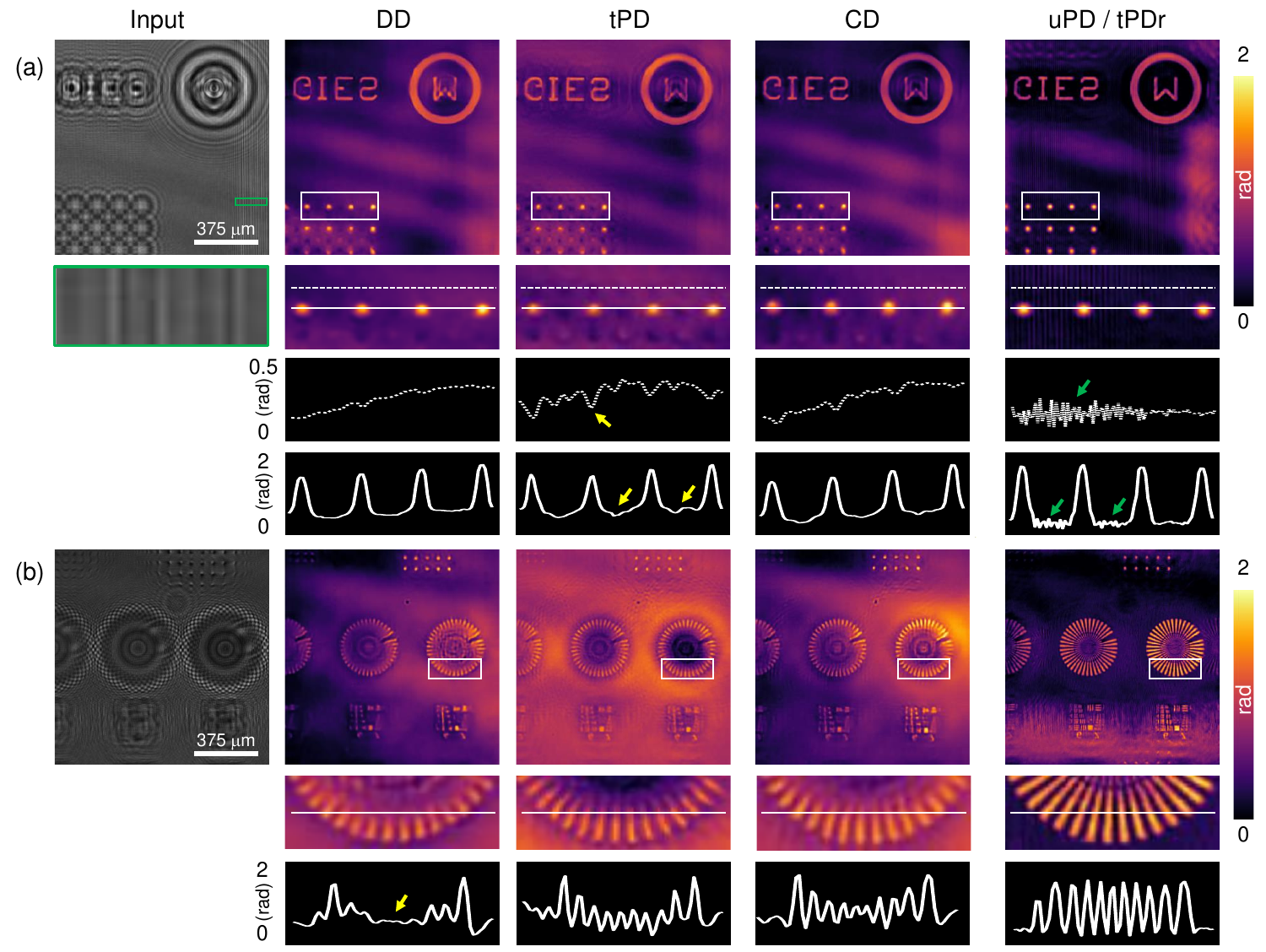}
\caption{Experimental tests of DD, tPD, CD, and uPD(tPDr). (a) inference results of one field of view. (b) inference results of another field of view.} \label{Fig:10_exp_test}
\end{figure}

We compare DD, tPD, CD, and uPD(tPDr) using experimental holograms with a defocus distance of 8.78mm from an open-source dataset of Ref.~\citeonline{gaoIterativeProjectionMeets2023a}. To match the defocus distance of the experimental hologram, we use ImageNet to generate corresponding datasets for the network training. Inference results of the standard phase object are given in Fig.~\ref{Fig:10_exp_test}. 

Overall, uPD and tPDr with multiple-cycle inferences have the best results, as seen from the neatly drawn peaks and valleys. It should be noted that due to the presence of redundant diffraction fringes at the edge of the hologram (see the green box in Fig.~\ref{Fig:10_exp_test}(a)), unwanted fluctuations appear in the background of the uPD and tPDr inference results (see the green arrows in Fig.~\ref{Fig:10_exp_test}(a)). Among the remaining one-time inference methods, the background fluctuations of the tPD results are larger (see the yellow arrows in Fig.~\ref{Fig:10_exp_test}(a)), while the detailed information of the DD results is weaker (see the yellow arrows in Fig.~\ref{Fig:10_exp_test}(b)). As a combination of DD and tPD, CD better considers detailed and background information. It should be noted that as the training dataset further expands, the neural network's accuracy will increase accordingly. In addition, we also test tissue slices and get similar conclusions, as detailed in Fig.~S2 of the Supplementary Material.

\section{Conclusion}

We introduced the principles of DD and PD strategies for deep learning phase recovery in the same context. On this basis, we compared the time consumption and accuracy of DD, uPD, tPD, and tPDr, and found that uPD and tPDr achieve the highest accuracy with multiple inferences, tPD prefers the high-frequency detailed phase while DD favors the low-frequency background phase. Therefore, we proposed CD to balance high- and low-frequency information. Furthermore, we found that tPD generalizes better than DD for the case of inferring dense samples using neural networks trained on sparse samples. As for the case of inferring phase and amplitude simultaneously, we revealed the reason why DD is stronger than tPD, that is, the dataset for DD implicitly contains the mapping relationship from holograms to phase and amplitude while tPD may encounter situations where multiple network outputs phases and amplitudes correspond to a same hologram. To alleviate the ill-posedness of tPD, we proposed solutions by aperture constraints or multiple hologram inputs. In addition, we used the case of imaging aberration to demonstrate that DD can learn more about the prior implicit in the dataset whereas PD can only learn the prior in numerical propagation. Finally, we verified with experimental data that uPD and tPDr have the highest accuracy and that CD balances high- and low-frequency information better than DD and tPD. 

We list some related papers with open-source code for readers to make further comparisons\cite{huangSelfsupervisedLearningHologram2023,rogalskiPhysicsdrivenUniversalTwinimage2024,yaoAutoPhaseNNUnsupervisedPhysicsaware2022,hoidnPhysicsConstrainedUnsupervised2023}.

\appendix    

\section{Dataset generation}
\label{app:data_gene}
Three publicly available image datasets (ImageNet, LFW, and MNIST) are used to generate phases and amplitudes, and then the corresponding holograms at a certain propagation distance are computed via numerical propagation. The training and testing datasets contain 10,000 and 100 data, respectively. The size of all data is set to $256\times 256$. The propagation distance is set to 20 mm and 8.78 mm for the simulation comparisons and the experimental tests, respectively. In the code, we provide a hyperparameter ``pad" to choose whether to use the way of ``padding and cropping" to eliminate edge diffraction effects (see Fig.~S3 of the Supplementary Material).

\section{Network implementation}
\label{app:net_opera_set}
The Adam optimizer with an initial learning rate of 0.001 is adopted to update the weights and biases. The Adam weight decay of uPD and tPDr is set to 0.001. The learning rate decreases to 0.95 of its current value every 5 or 10 epochs until it approaches 0.00001. The batch size of DD, tPD and CD is set to 16. The neural network training epoch of DD and PD is set to 100. The inference cycles of uPD and tPDr are set to 10,000 and 1000, respectively. All the neural networks are based on Pytorch (2.0.0) with Python (3.8.18). All operators run on a compute server equipped with AMD Ryzen Threadripper PRO 3955WX and NVIDIA GeForce RTX 3090. 

\subsection* {Code and Data, and Materials Availability} 
Code and data are available at \href{https://github.com/kqwang/DLPR}{https://github.com/kqwang/DLPR}.


\bibliography{article}   
\bibliographystyle{article}   

\end{spacing}

\title{Supplementary Material for \\
Deep learning phase recovery: data-driven, physics-driven, or combining both?}

\renewcommand{\cftdotsep}{\cftnodots}
\cftpagenumbersoff{figure}
\cftpagenumbersoff{table} 
\maketitle

{\noindent \footnotesize\textbf{*}Kaiqiang Wang,  \linkable{kqwang.optics@gmail.com}; Edmund Y. Lam,  \linkable{elam@eee.hku.hk} }

\begin{spacing}{2}   

To explore the tolerance of methods to defocus distances, we generate holograms with propagation distances from 15mm to 25mm and infer them using neural networks trained on the 20mm dataset. The SSIM of the inference results and the corresponding samples are shown in Fig.~S1. It can be seen that DD is more tolerant than tPD, while CD neutralizes them.

\renewcommand\thefigure{S\arabic{figure}}
\setcounter{figure}{0}
\begin{figure}[!ht]
\centering
\includegraphics[width=1\linewidth]{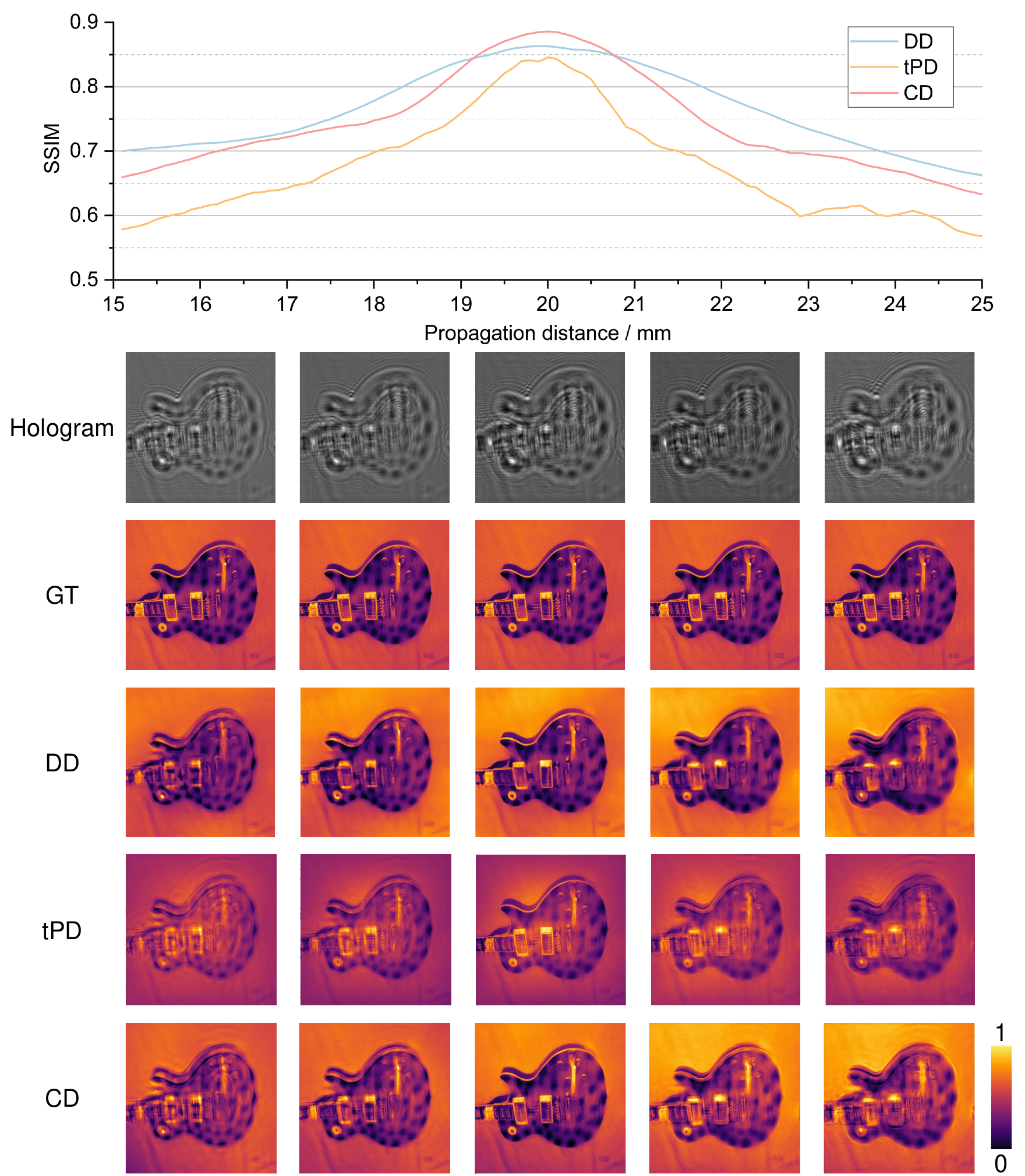}
\caption{Defocus distance tolerance tests of DD, tPD, and CD.}\label{Fig:S1_defocus}
\end{figure}

We test all methods using holograms of tissue slices. The results are shown in Fig.~S2. Compared with DD, as a joint strategy of data and physics, CD also infers more high-frequency information like tPD, see the yellow arrow of Fig.~S2. Since the inferences go through multiple cycles, the results of uPD and tPDr contain richer information, see the green arrow of Fig.~S2.

\begin{figure}[!ht]
\centering
\includegraphics[width=0.8\linewidth]{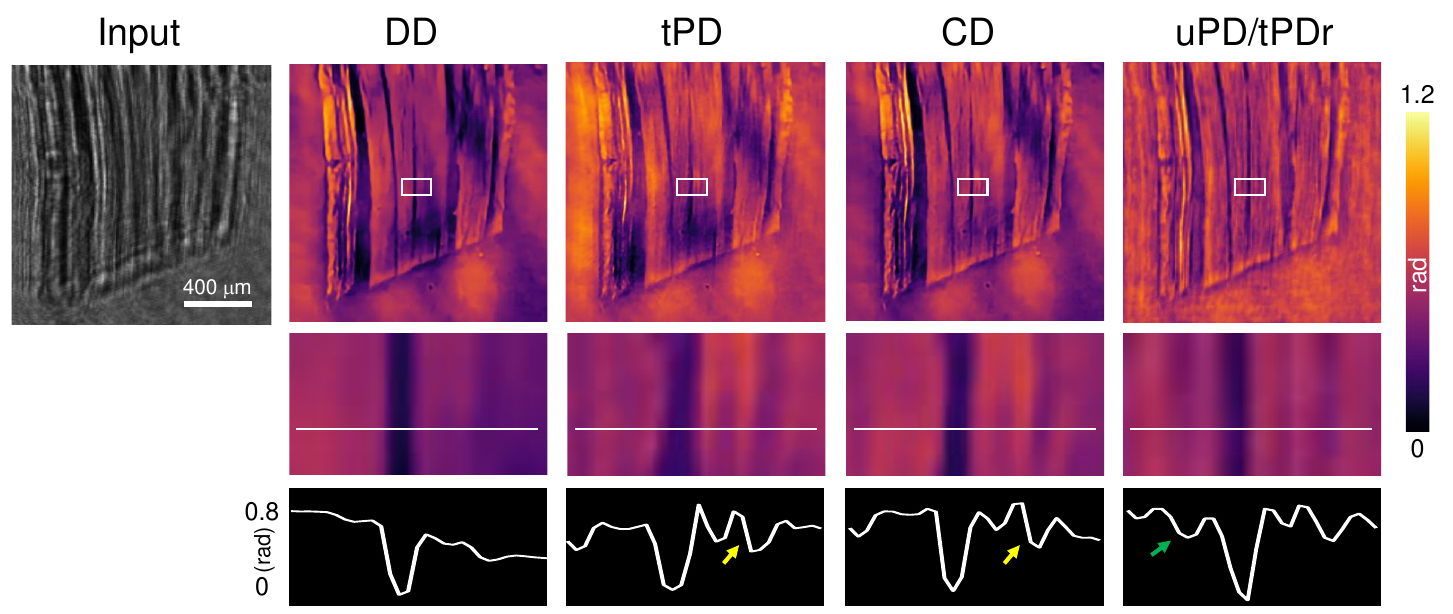}
\caption{Experimental tests of DD, tPD, CD, and uPD(tPDr) for tissue slices.}\label{Fig:S2_exp_tissue}
\end{figure}

We provide a hyperparameter ``pad" in the code to eliminate edge diffraction effects through padding and cropping. As shown in Fig.~S3, a hologram is generated directly through numerical propagation (the upper part), or ``padding and cropping" is added to eliminate edge diffraction effects (the lower part).   

\begin{figure}[!ht]
\centering
\includegraphics[width=0.8\linewidth]{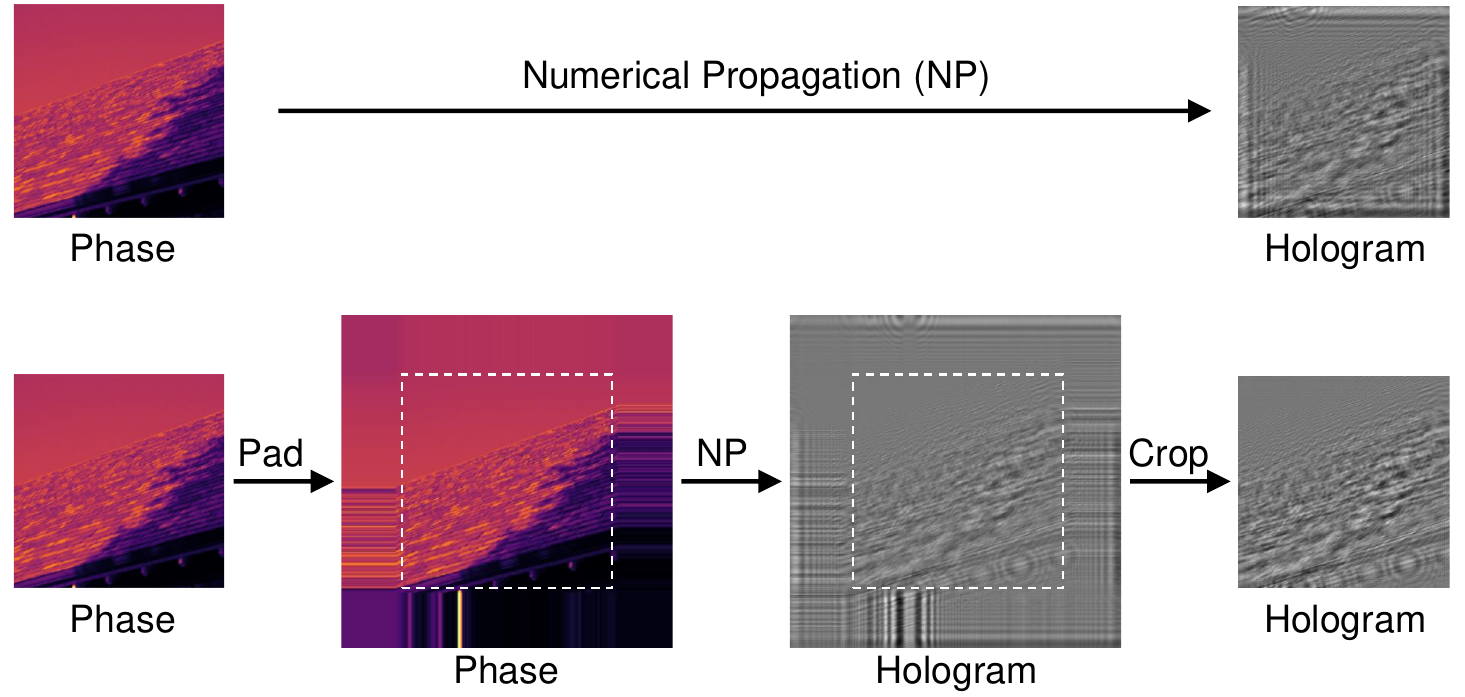}
\caption{An example of hologram generation. Generate holograms directly by numerical propagation (upper part) or use ``padding and cropping" to eliminate edge diffraction effects (lower part).}\label{Fig:S3_holo_gene}
\end{figure}

\end{spacing}
\end{document}